\documentclass[preprint]{aastex631}

\pdfoutput=1 
\usepackage{amsmath,amstext}
\usepackage[T1]{fontenc}
\usepackage{gensymb}
\usepackage{subfigure}
\usepackage{longtable}

\usepackage{subfigure}
\usepackage{savesym}
\savesymbol{tablenum}
\usepackage{siunitx}
\restoresymbol{SIX}{tablenum}
\DeclareSIUnit{\year}{yr}
\DeclareSIUnit{\mas}{mas}
\DeclareSIUnit{\astronomicalunit}{au}
\DeclareSIUnit{\parsec}{pc}
\newcommand\redsout{\bgroup\markoverwith{\textcolor{red}{\rule[0.5ex]{2pt}{0.4pt}}}\ULon}

\shorttitle{Speckle Imaging of Low-Mass Disk Wide Binaries}
\shortauthors{Hartman et al.}
\graphicspath{{./}{figures/}}

\begin{document}

\title{{Resolving the Unresolved: Using NESSI to Search for Unresolved Companions in Low-mass Disk Wide Binaries}}

\correspondingauthor{Zachary Hartman}
\email{zachary.hartman366@gmail.com}
\author[0000-0003-4236-6927]{Zachary D. Hartman}
\affiliation{NASA Ames Research Center, Moffett field, CA 94035, USA}
\affiliation{International Gemini Observatory/ NSF NOIRLab, 670 A'ohoku Place, Hilo, HI 96720, USA}

\author[0000-0002-8552-158X]{Gerard van Belle}
\affiliation{Lowell Observatory, 1400 West Mars Hill Road, Flagstaff, AZ 86001, USA}

\author[0000-0002-2437-2947]{S\'{e}bastien L\'{e}pine}
\affiliation{Georgia State University,Department of Physics \& Astronomy, Georgia State University, 25 Park Place South, Suite 605, Atlanta GA, 30303, USA}

\author[0000-0002-0885-7215]{Mark E. Everett}
\affiliation{NSF NOIRLab, 950 N. Cherry Ave., Tucson, AZ 85719, USA}

\author[0000-0003-3410-5794]{Ilija Medan}
\affiliation{Department of Physics and Astronomy, Vanderbilt University, Nashville, TN 37235, USA}



\begin{abstract}

Stellar systems consisting of three or more stars are not an uncommon occurrence in the Galaxy. 
Nearly 50\% of solar-type wide binaries with separations $>1000$ au are actually higher-order multiples with one component being a close binary. 
Additionally, the higher-order multiplicity fraction appears to be correlated with the physical separation of the widest component.
These facts have motivated some of our current theories behind how the widest stellar systems formed, which can have separations on the order of or larger than protostellar cores. 
However, it is unclear if the correlation between wide binary separation and higher-order multiplicity extends to low-mass binaries. 
We present initial results of an ongoing speckle imaging survey of nearby low-mass wide binaries. 
We find an overall higher-order multiplicity fraction for our sample of $42.0\% \pm 10.9\%$. 
If we include systems where \textit{Gaia} indicates that a companion is likely present, this fraction increases to $62.0\% \pm 14.2\%$. 
This is consistent with previous results from both higher-mass stars and a previous result for low-mass wide binaries. 
However, we do not detect the expected increase in higher-order multiplicity fraction with separation, as was seen with previous studies. 
We briefly explore why higher-order multiplicity statistics could be different in low-mass stars, and what the significance might be for models of wide binary formation.

\end{abstract}


\section{Introduction}
Wide stellar binary systems with separations larger than 1000 au are some of the most important astrophysical laboratories for astronomers. 
These systems are typically found as common proper motion pairs through statistical methods by comparing different astrometric parameters of stars, such as their positions, proper motions, parallaxes, and radial velocities, and finding pairs in relative proximity that share common space motion \citep{2004chaname,2007lepine,2010dhital,2011lepine,2011shaya,2012tokovinin,2014tokovinina,2015dhital,2016deacon}. 
With the release of the \textit{Gaia} data, the number of known wide binaries has dramatically increased in recent years \citep{2017andrews,2017oh,2017oelkers,2018elbadry,2018coronado,2019jimenezesteban,2020hartman,2020tian,2021elbadry}. 
For stellar astronomers in particular, these systems are useful tools for examining the properties of the component stars, notably because the components can be seen as two non-interacting ``single'' stars that are coeval and chemically homogeneous, like a mini star cluster \citep{2020hawkins,2021nelson,2024lim}. 
Because of this, these systems are commonly used as calibrators for different scaling relations, such as gyrochronology relations \citep{2012chaname,2017janes,2018godoyrivera,2023silvabeyer} and metallicity relations \citep{2007lepinemetal,2013mann,2014newton,2014mann,2017veyette,2019andrews,2020wheeler}. 
Furthermore, wide binaries with white dwarfs are commonly used to obtain age estimates for the white dwarfs \citep{2024barrientos,2024heintz}.

Even with their relative importance to stellar astronomers, there are still open questions regarding the widest binary systems. 
In particular, our knowledge about the physical properties and statistics of these systems remains fragmentary, as highlighted in \citet{2023offner}. 
This, in turn, has left the formation mechanism for the widest systems as an unsolved problem. 
Numerous scenarios have been proposed to explain how systems with separations larger than that of a protostellar cloud core can form in a star-forming environment. 
Some of these theories include the unfolding of triple systems \citep{2012reipurth}, the binding of stars during the cluster dissolution phase \citep{2010Kouwenhoven,2010moeckel}, the binding of adjacent protostellar cores in star forming regions \citep{2017tokovinin}, and turbulent fragmentation \citep{2014bate}. 
Recent work from \citet{2022hwang} showed that the eccentricity distribution for wide binaries with separations larger than 1000 au is super-thermal, which matches predictions from both the turbulent fragmentation and the unfolding of triple systems scenarios. 
These two formation models also predict that the eccentricity should increase with larger separations.
However, \citet{2022hwang} found that the eccentricity distribution remained uniform up to $10^{4.5}$ au, which disagrees with the two scenarios.
They do note that eccentricity measurements are less reliable at larger separations, and that other factors, such as gravitational interactions with the Galactic environment, may have an unknown impact on the orbits of large separation binaries. 
Regardless, \citet{2022hwang} highlights that wide binary formation is likely the combination of a variety of different formation scenarios rather than just a single dominant channel.

In order to further study how the widest systems form, one key property that must be determined across stellar mass is the higher-order multiplicity fraction of these systems, i.e., how many binaries are actually triples, quadruples, etc.
In particular, a major difference between the turbulent fragmentation and unfolding scenarios is that the unfolding scenario predicts that the vast majority of systems will be higher-order multiples, while the turbulent fragmentation scenario does not. 
Therefore, measuring the frequency of higher-order multiples may give insights into how prevalent one scenario is over the other.
While common belief is that wide binaries usually consist of two ``single'' stars, this idea has been consistently challenged over the past couple of decades. 
Over half of wide solar-type binaries are found to actually be higher-order multiples with this fraction increasing as a function of projected physical separation \citep{2010raghavan,2014tokovininb,2019moe,2023offner}. 
This has also been observed for higher mass binaries \citep{2017moe}. 
However, most studies have focused on higher mass systems as they are bright and make easier targets to search for close companions that are unresolved by large astrometric surveys like \textit{Gaia}. 

\citet{2010law} is one exception. 
In this study, they observed 36 M+M wide binaries and found that $45\%\substack{+18\% \\ -16\%}$ are higher-order multiples, after bias-correcting their analysis. 
This bias-correction took into account selection bias from using the SDSS DR7 photometric catalog, the suppression of triple systems with an unresolved companion, narrow-angle incompleteness, and contrast-ratio incompleteness. 
They also found evidence of an increase of higher-order multiplicity as a function of projected physical separation.
However, \citet{2010law} had large uncertainties on their results and only observed wide binaries with separations less than 6500 au. 
Since then, there have been few studies directly examining the higher-order multiplicity of low-mass binaries. 
In a previous paper, we examined a sample of K+K wide binaries and, exploiting the fact that if a wide binary is a true wide binary it should fall on the same metallicity track, we searched for over-luminous components that may hide an unresolved companion \citep{2020hartman,2022hartman}. 
We found the lower limit on the higher-order multiplicity of this sample to be 40\%, however, our study was focused only on K+K wide binaries and was primarily sensitive to companions with close to equal-mass ratios. 

In this paper, we expand upon the study of \citet{2010law} and include systems at larger physical separations to see if the higher-order multiplicity of low-mass binaries does increase with increasing separation. 
In Section 2, we describe the sample selection and speckle imaging observations.
In Section 3, we present the results of our observations and combine them with the results from \citet{2010law}.
Finally, in Section 4, we discuss our results with an eye towards how these systems formed. 

\section{Sample Selection and Observations}
\subsection{Selection of the Wide Binary Sample}
The sample for this paper was taken from the SUPERWIDE catalog of wide binaries. This catalog was created using the second \textit{Gaia} data release supplemented with high-proper motion stars from the SUPERBLINK catalog (DR2, \citet{2018gaiadr2,2018Lindegren}).
The method used to identify the wide binaries is detailed in \citet{2020hartman}.
We provide a brief summary of how the SUPERWIDE catalog was created below. 
Since SUPERWIDE was published, \textit{Gaia} DR3 has been released \citep{2023gaiadr3}.
We have matched the DR2 identifiers in SUPERWIDE to their corresponding DR3 identifiers using the method of \citet{2021medan}. 
For the present paper, all data used in calculations is the DR3 data, which caused some of the targets to fall a little outside of the DR2 sample selection.
For completeness, we have decided to keep these targets in our analysis.

Starting with the complete set of $\sim5.9$ million high proper motion stars ($>$\SI{40}{\mas\per\year}) in \textit{Gaia} DR2 supplemented by a modest number of additional high proper motion stars from the SUPERBLINK catalog that are real but not listed in DR2, we conducted a two stage Bayesian analysis that calculated the probability of any two stars to be physical pairs (as opposed to chance alignments) based on their angular separations, proper motion differences, and distance/parallax differences. 
The first stage takes the angular separations and proper motion differences and uses empirical model distributions of these two parameters for both chance alignments and real binaries to calculate a first pass real-binary probability. 
We then select only pairs with a first-pass probability greater than $10$\% and with a parallax error on both components less than $10\%$.
We run another Bayesian analysis using just the difference in the distances between the components in the selected pairs, again using empirical model distributions for real binaries and chance alignments. 
The end result is the SUPERWIDE catalog listing 99,203 high proper motion pairs with probabilities of being gravitationally bound $>95\%$. 

From this catalog, we make the following cuts to select our sample of targets.
We require the probability from SUPERWIDE to be $> 90\%$ to ensure that our pairs are most likely to be bound systems.
We adopt the slightly lower value of $90\%$ rather than $95\%$ to take into account that components of higher-order systems may register larger differences in their measured proper motions and/or parallaxes due to hidden 3rd component perturbations. 
To select low-mass systems, we require both components to have $G_{BP} - G_{RP} > 1.01$, and $M_G > 4$. 
We also require $M_G$ to fall above a linear demarcation in the H-R diagram which separates the white dwarfs from the main sequence, defined to be:

\begin{equation*}
M_G \leq 3 * (G_{BP}-G_{RP}) + 4.25
\end{equation*}

We require the primary's distance to be $< 100$ pc and the pair to have a projected physical separation, $\rho > 1000$ au. 
To ensure our sample includes primarily stars from the disk population, we add a tangential velocity cut $< 120 \SI{}{\kilo\metre\per\second}$. 
We further select only pairs where both the primary and secondary components have $G<15$ so both components are observable with the telescope used in this survey. 

These cuts create a sample of 1650 wide system candidates. 
It was noticed after the fact that 308 of these systems have resolved third or fourth components in Gaia DR3, with roughly half of them as higher-order multiples with a higher mass companion or a white dwarf.
We will not include any higher-order multiples with a higher-mass companion or white dwarf in our analysis as our primary goal is to determine the higher-order multiplicity fraction of low-mass systems; our observations of these higher-mass systems will however be listed as appropriate. 
It was also later noticed that our cut on $G_{BP}-G_{RP}$ failed if a source in \textit{Gaia} did not have a $G_{BP}$ or $G_{RP}$ value, but it had astrometric and \textit{Gaia} $G$ band photometry.
Our selection code kept them in our list at this point. 
For the single component that this occurred to and was observed with NESSI (\textit{Gaia} DR3 1632110529080430592, Binary 14), it is marked in the relevant table with a placeholder for the $G_{BP}-G_{RP}$ value.
We will talk about this component more later on as it is part of a \textit{Gaia} resolved triple system along with several other similar systems.

\subsection{NESSI Speckle Observations}

Observations using the NN-Explore Exoplanet Stellar Speckle Imager (NESSI, \citet{2018PASPScott,2021Howell}) on the 3.5m WIYN telescope were carried out in queue mode over a total of 4 nights in the 2019A semester. 
NESSI is an efficient instrument capable of observing hundreds of targets per night depending on the magnitudes of the targets being observed. 
Light is split by a dichroic centered at 686.4nm and allows for the target to be observed using blue and red filters. 
For our survey, we collected data using two narrow-band filters ($\Delta\lambda =$ 40 nm) centered on 562 nm and 832 nm. 
We will refer to these filters as the ``blue'' and ``red'' filters throughout the rest of the paper.

56 wide systems were observed over the program's lifetime plus the primary component of another system. 
For this paper, we will report the observations and results for all systems, however, we will remove the system where only one component was observed from the later analysis (Binary 46). 
In each case, only two stars were examined even if the system was a resolved triple in \textit{Gaia}. 
This was due to the resolved third component miscue talked about in the previous section. 
In some cases, the resolved third component was close enough to one of the observed components that it was resolved. 

Targets were selected from the 1650 systems at random. 
Table \ref{tab:targets} provides the \textit{Gaia} astrometric and some photometric information on the observed targets. 
For each examined pair of stars, the primary is put on one line and then the secondary is the next line. 
We provide the Binary ID we assigned to the pair from 1 to 57, R.A., Decl., $\mu_{R.A.}$, $\mu_{Decl.}$, Parallax, $G$ magnitude, $G_{BP} - G_{RP}$, Gaia DR3 ID, and the observation date for the target with NESSI. 
The ordering of the pairs is kept the same across all relevant tables.

For speckle imaging with NESSI, the typical observing strategy is to observe in data cubes of 1000 frames, with each frame being taken using a 40 ms exposure. 
For each target, a certain number of ``cubes'' are observed depending on the magnitude of the target. 
During our observations, this number was selected by the observer on-site who would know the current conditions and adapt to take the appropriate number of ``cubes'' per target.
For the brightest targets, one to three cubes are typically collected, while for the faintest targets, ten to twenty cubes may be needed.
Along with each target, a single cube of data is collected on a known nearby single star which is used as point source standard. 
For our targets, we only observed one point source standard per target system as these systems are relatively close on the sky ($\rho < 1\degree$) and the average time to complete a pair plus the standard was 19 minutes.
Furthermore, calibration binaries were observed throughout the run to get a measure of the pixel scale and image orientation of the detectors.
For our observations, the pixel scales were measured as 0.0182829 \arcsec/pixel in the blue filter and 0.0183121 \arcsec/pixel in the red filter.

\begin{figure}
\centering
\subfigure{
\includegraphics[width=0.48\textwidth]{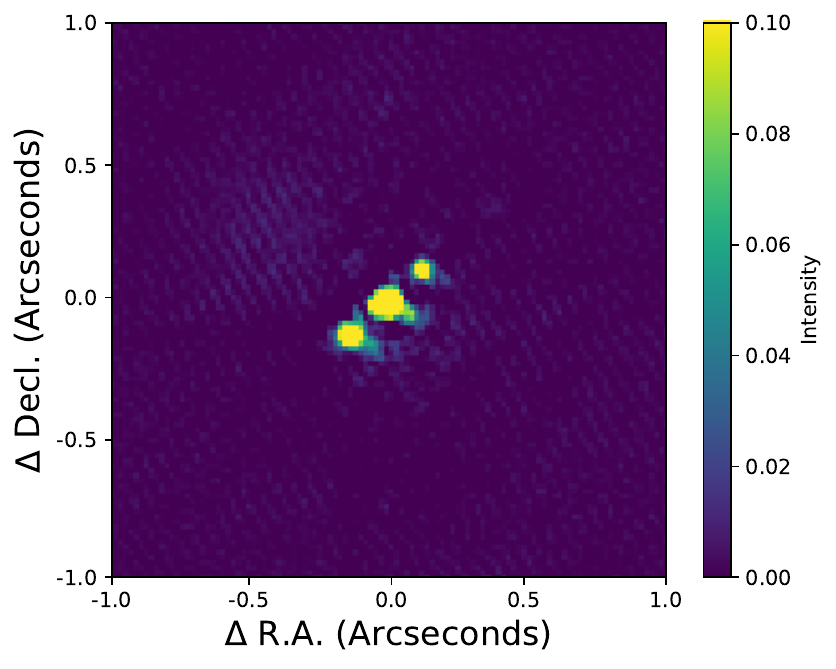}}
\subfigure{
\includegraphics[width=0.48\textwidth]{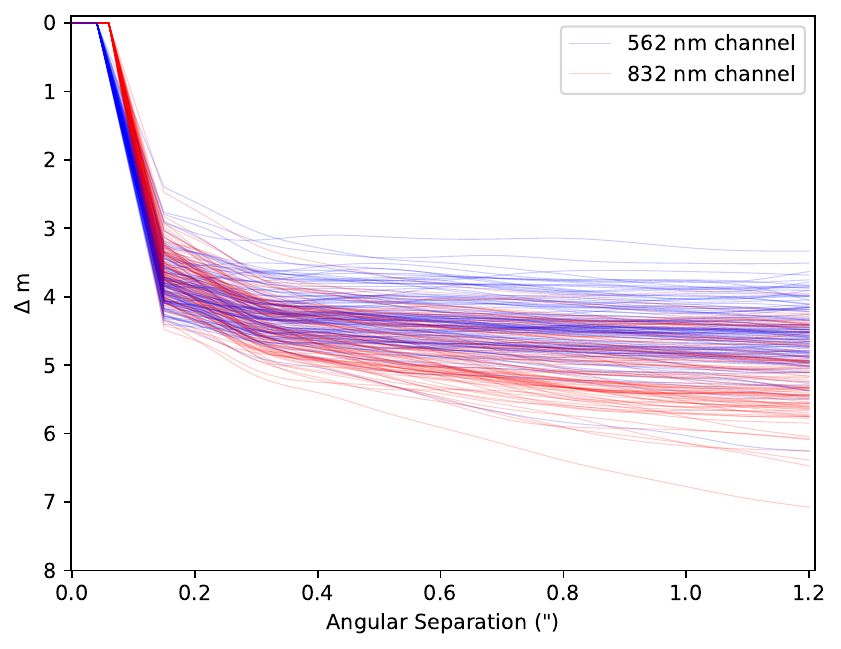}}
\caption{\label{fig:NESSIobs}Data products produced from the speckle data reduction. Left Panel: Reconstructed image in the red filter (832 nm) for the target star (UCAC4706-053644). A binary is detected at 0.172\arcsec with a delta magnitude of 0.52. A ghost image at 180\degree is seen as well. This ghost image is not a third companion but is an artifact of the image reconstruction. The true companion is the brighter component to the lower left of the target star in the middle. 
Right Panel: Contrast curves for all of the observed targets in our sample. Blue lines highlight the contrast curves for the 562 nm filter, while red lines do the same for the 832 nm filter.}
\end{figure}

The data was reduced by Co-I Everett using a speckle reduction pipeline which is explained in more detail in  \citet{2009Horch,2011Horcha,2011Horchb,2011Howell}. 
Briefly, this pipeline uses a Fourier analysis to compute the angular separation ($\rho$), position angle ($\theta$), and delta magnitudes ($\Delta m$) for any detected companions. 
The pipeline then uses a bi-spectrum analysis to reconstruct the image and produce a contrast curve that is measured from the image \citep{1983Lohmann}. 
This curve is calculated by examining the maximum and minimum values within concentric annuli centered on the bright target and determining the $5\sigma$ deviation from the mean in each annuli. Example outputs from this pipeline are shown in Figure \ref{fig:NESSIobs}. 
The left plot shows the reconstructed image in the red (832 nm) filter for one of our target stars (UCAC4706-053644).
The central point is the target star and there appears to be two companions to the star. However, the apparently fainter companion is actually a ghost image and not a true third companion. 
It is rather an artifact of the Fourier analysis. 
The real companion is the brighter object located in the lower left.
The right plot shows the contrast curves for all of our observed targets.
If a companion lies in the regions above these curves, then we would expect to detect it. 
As seen, we reach an average $\Delta m$ of 4 for our targets.
It extends out to 1.2\arcsec, as beyond this point, the speckle patterns can begin to become uncorrelated.
 
\subsection{Crossmatch with Simbad}
We also crossmatch our sample with Simbad and search for previous detections. 
The results of this search are shown in Table \ref{tab:nessisimbad}, where we note any sources that have conducted observations searching for multiplicity in either component of the binary. 
The columns are the Binary ID given in this study, the Washington Double Star (WDS) Catalog ID if available, the \textit{Gaia} DR3 ID for each component, whether the component has been studied previously along with the method used (RV - Radial Velocities, TS - Transit Searches, PMA - Proper Motion Anomaly/\textit{Gaia}-Hipparcos accelerations, Astro - \textit{Gaia} astrometric searches, Accel - \textit{Gaia} acceleration searches, and HRI - High-Resolution Imaging), and the result of that search (Column Is \# Binary?).
In some cases, the stars are not marked on Simbad as being part of the \textit{Gaia} Non-Single Star catalog (NSS, \cite{2023gaiastellarmult}). 
We mark these components here even if Simbad does not.
This primarily impacts the \textit{Gaia} astrometric and acceleration searches.

We find a number of targets have been observed previously using a variety of different methods. 
If the target is observed with different methods, each method is listed on a separate line. 
In cases where the target has been observed more than once with the same method, we chose the paper which is potentially best suited for detecting companions. 
This largely impacted those targets that were examined using the proper motion anomaly and \textit{Gaia}-Hipparcos accelerations as they were studied in both \citet{2019kervella} and \citet{2021brandt}.
We choose to cite \citet{2021brandt} in the table due to the fact that this paper includes \textit{Gaia} EDR3 rather than just DR2 and we make a determination of whether to mark it as detecting a binary or not if the $\chi^2$ value is above 11.8 or their $3\sigma$ detection limit. 
For the two components which are marked as possible binaries, their $\chi^2$ values are near or above 1000.

For the high resolution imaging sources, we chose the studies that have the highest resolution to cite in the table. 
For RV studies, we report only those that specifically target the stars looking for either exoplanets or binaries.
Some of the targets have multiple RVs from separate studies (like APOGEE and RAVE) and we leave an analysis on those stars for a later paper. 
Similarly for transit searches, most, if not all, of these stars have Kepler and/or TESS data available with several stars targeted in studies looking at the rotation of the stars. 
However, it was unclear from those studies if a binary was detected.
Therefore, unless a study specifically points out that the component was discovered to be an eclipsing binary, we leave reanalysis of the available Kepler and TESS data for a later paper. 
The astrometric and acceleration searches are used for systems that have been flagged by the \textit{Gaia} NSS as having either an astrometric binary or indications of binary based on acceleration.

In total, we find that 33 of the wide binary components in our sample have been previously observed as part of RV, TS, PMA, Astro., Accel., or HRI studies. 
Two stars show signs of multiplicity from \textit{Gaia}-Hipparcos accelerations, one star is an eclipsing binary, two stars are resolved astrometric binaries, two are astrometric binaries from the \textit{Gaia} NSS, and four stars are spectroscopic binaries. 
We will comment on individual targets in the next section.

\section{Results}
Our NESSI observations resolve nineteen companions to the components of the 57 wide systems examined. 
Table \ref{tab:nessiresults} provides the \textit{Gaia} DR3 source id for both components, the angular separation ($\rho$), position angle ($\theta$), and $\Delta$m if a companion is detected around either component in the 562nm (blue, b in the table) or 832nm (red, r in the table) filters, whether the system is a \textit{Gaia} resolved triple (Y/N), the SUPERWIDE probability, and the projected physical separation of the wide binary that was examined. 
In some cases, the close companion was only detected in the red filter.
In the case where no close companion is detected in the speckle observations, the columns have a placeholder. 
In most instances, the red and the blue results matched with one exception which we discuss below.

Eight of the nineteen are close companions that are not resolved by \textit{Gaia} and are in systems where the \textit{Gaia} DR2 catalog only listed two wide co-moving sources (Binaries 1, 10, 15, 16, 26, 27, 32, 36). 
Another five are triples that were resolved in \textit{Gaia}, with the \textit{Gaia} catalog already listing all three components (Binaries 12, 13, 29, 48, 55). 
We have resolved the closer two in the presented observations. 
The remaining six companions are members of new quadruple systems, which are discussed. 

\subsection{Notes on Selected Systems}

\subsubsection{Binary 11}
We report no new speckle detections in Binary 11. 
However, an examination of the literature reveals that this system is a potential quadruple. 
1371592106557884928 is reported as a spectroscopic binary in both \citet{2019Sperauskas} and \citet{2023gaiastellarmult}.
It was also in the PMA analysis of \citet{2021brandt} with a low value of 4.054 for the $\chi^2$ value.
On the other hand, 1371779401491517952 had an extremely large $\chi^2$ value of 4016 from the PMA analysis.
Therefore, we consider this system as a higher-order multiple and as a potential quadruple system.

\subsubsection{Binary 14}
Binary 14 is a \textit{Gaia} resolved triple with the third \textit{Gaia} companion close to 1632110529080430592, which we find with our speckle imaging observations. 
However, we also report the discovery of potentially a fourth companion in the system around 1632112552014817024, making this system a potential quadruple system.
1632110529080430592 also had no $G_{BP}-G_{RP}$ value, however, it's magnitude is near the other components.
Therefore, we will assume it is a low-mass system for the later analysis.

\subsubsection{Binary 25}\label{sec:bin25}
It was discovered that component \textit{Gaia} DR3 1449301838900762880 of Binary 25 was a triple system itself, making the entire system a 3+1 quadruple.
In examining the \textit{Gaia} data for this star, it was found that there was a nearby object at roughly the same separation and delta magnitude as one of the resolved components (\textit{Gaia} DR3 1449301843195744768). 
This object does not appear in the SUPERWIDE catalog most likely because the proper motions are very different between the object and the wide binary components, $\sim10$ mas $\SI{}{\per\second}$ different. 
Furthermore, the RV measurements are also very different, $45 \SI{}{\kilo\metre\per\second}$ compared to $6 \SI{}{\kilo\metre\per\second}$. 
However, this could be because the object itself is a close binary, leading to \textit{Gaia} detecting the orbital motion of the pair in the RV measurements. 
To double check whether the system is bound, we apply the criteria for selecting systems from \citet{2023Tokovinin}, which is more relaxed towards triple systems than the analysis used to construct the SUPERWIDE catalog. 
This method identifies wide binaries from the following requirements:

1.) Parallax difference $<$ 1 mas

2.) Projected physical separations $<$ 20,000 au

3.) Relative projected speed $< 10 * (10^3 / s)^{0.5}$, where s is the projected physical separation

We find that 1449301843195744768 satisfies the conditions to be bound to the closer companion (1449301838900762880). 
However, it barely fails the criteria for being bound to the wide companion (1449467422775745792) by $0.42\SI{}{\kilo\metre\per\second}$ in relative projected speed. 
The error on the projected speed is $0.015\SI{}{\kilo\metre\per\second}$
Because it seems bound to one of the components, we decided to keep this system as a higher-order multiple in our analysis with the caveat that it may be proven to be unbound in the future.
Because the third and fourth components are not in SUPERWIDE, we do not consider this to be a \textit{Gaia} resolved triple however. 
In Table \ref{tab:nessiresults}, we also add in 1449301843195744768 as part of Binary 25 in order to portray that it is a binary itself.

\subsubsection{Binary 27}
We report a speckle detection in Binary 27 around component 1595017404806620928.
However, \textit{Gaia} also reports that the other component, 1595020325384386304, is a binary with a period of 625.1 days \citep{2023gaiastellarmult}.
This makes Binary 27 a possible quadruple system.

\subsubsection{Binary 29}
Binary 29 is another \textit{Gaia} resolved triple with the third companion located $\sim2$\arcsec away from DR3 389658300887082496.
However, the wide component, 388149427335376256, was flagged by \textit{Gaia} as an astrometric binary with a period of 58.7 days \citep{2023gaiastellarmult}.
This makes Binary 29 a possible quadruple system.

\subsubsection{Binary 32}
Binary 32 was resolved as a triple system by our speckle observations with a close companion located 0.351\arcsec away from \textit{Gaia} DR3 1777019255313788928. 
However, this component is also listed as a potential single-lined spectroscopic binary in \textit{Gaia} DR3 \citep{2023gaiastellarmult}.
It is listed as having a period of 15 days, well below our detection limit with NESSI. 
We count this as another possible quadruple system.

\subsubsection{Binary 33}
Binary 33 is also a resolved \textit{Gaia} triple with the third component around 2829469414599661568 separated by 2.47\arcsec.
Additionally, 2829469414599661568 has a close companion detected at 0.086\arcsec\,in the red filter and is very close in magnitude to the primary star. 
Between the red and the blue results, the position angle is flipped 180\degree. 
It is uncertain which orientation is correct so we leave the results as is. 
Interestingly, \textit{Gaia} also reports 2829469414599661568 as a binary. 
The period listed is 138.9 days \citep{2023gaiastellarmult}.
Taking the separation of our speckle detection as the semi-major axis and assuming a circular orbit and the mass of the system to be 1 solar mass, we estimate the period for that component to be 14.3 years. 
This implies that our speckle detection and the detection from \textit{Gaia} are two different detections, meaning that this system is a potential quintuple system assuming all of these components are verified as bound companions.
We do note, however, that this system is not a low-mass multiple system as 2829469414599661440 falls below our white dwarf cutoff line.

\subsubsection{Binary 42}
Binary 42 was reported as a potential triple system by the NESSI observations. 
Another star was seen at 2.8\arcsec away from \textit{Gaia} DR3 2231300248317642112. 
However, SUPERWIDE does not contain a match to this star and, upon checking the \textit{Gaia} archive, it was discovered that there were no stars around the component with matching proper motions and parallaxes. 
Therefore, we determined this detection to be of a chance alignment rather than a true binary and we removed it from our results and count the system as a wide binary for our analysis.

\subsection{Checking Astrometric Precision}
In order to measure the precision of our measured separations and position angles, we use the method of several previous papers which have used NESSI data \citep{2024clark,2021colton,2021horch,2017Horch}. 
As NESSI provides observations in two bands, we can compute the difference between the blue and red channels to examine our precision.
In some cases, the companion is only detected in the red channel. 
We do not use those observations for this calculation. 
We also do not use binary 33 in this analysis as it has the 180\degree ambiguity problem.
The average value for the angular separation differences is 1.1 mas with a standard deviation of 10.7 mas.
These values are in line with previous studies using NESSI.
We do a similar exercise with the position angles, leading to an average of 1.11\degree and a standard deviation of 0.57\degree. 
These results are reduced by a factor of $\sqrt{2}$ as these results are being measured from two independent observations that should have the same uncertainty. Our final uncertainties in a single measurement are 7.56 mas for the angular separation and 0.40\degree for the position angles. 
These values are a little larger than previous results, although they are similar to \citet{2024clark}, who hypothesized that this could be due to the faintness of their sample. 
Our sample is also relatively faint compared to the pervious studies, which could indicate that this a factor here as well. 

\subsection{Comparison with Literature Results}
In comparing our speckle results with previous high resolution imaging results from Simbad, we find that most of our observations agree with those of past studies with two notable exceptions: 1632112552014817024 (Binary 14) and 1777019255313788928 (Binary 32).
Both were observed by Robo-AO in \citet{2020lamman} and \citet{2021salama} respectively, which were not reported to have companions at the times but for which we now have positive detections. 
For 1777019255313788928, we find that the companion has a $\Delta m$ of 1.96 in the red 832 nm filter and a separation of 0.351\arcsec. 
Given the contrast limits on this target in \citet{2021salama} of 1.5 in the i' band at 0.5\arcsec, we believe this target fell right below their detection limit. 
Our confidence that this is a true detection is also bolstered by the \textit{Gaia} Reduced Unit Weight Error (RUWE) and ipd\_frac\_multi\_peak (IPDfmp) values, which are very high at 9.85 and 34. 
RUWE is a measure of how good the \textit{Gaia} astrometric fit is to a single star model.
If it is close to one, then the star fits the model well. 
For those with higher values, it is potentially indicative of a binary companion which is interfering with the fit. 
IPDfmp is the fraction of double transits, where the \textit{Gaia} data for a certain window shows a double peak, indicating the detection of another companion. 
For 1632112552014817024, \citet{2020lamman} provides no information on the contrast limits achieved.
We find the companion in this system at a separation of roughly 0.23\arcsec with a $\Delta m$ of 0.45 in the red filter. 
We count this is another true detection as the \textit{Gaia} RUWE value is once again very high, 9.27.

\subsection{\textit{Gaia} Resolved Systems}
Table \ref{tab:nessitriple} gives the \textit{Gaia} DR3 source IDs for the systems that are higher-order multiples with three or more components that are already listed in \textit{Gaia} DR3, including those where the third component is not detected in our observations due to their separations from the targeted components, and whether all members are low-mass stars. 
Our criterion to identify the multiple systems as ``low-mass'' is for all components to have $G_{BP} - G_{RP} > 1.0$, $M_G > 4.$, and not be white dwarfs. 
As stated previously, we wish to only examine low-mass systems (K- and M-dwarf binaries) in this analysis. 
Therefore, any system which has a higher-mass star or a white dwarf is removed. 
The removed multiple systems are listed in Table \ref{tab:nessitriple} as those with N in the ``Low-mass Triple'' column. 
In four cases (Binaries 13, 14, 43, and 48), one of the components in the system does not have \textit{Gaia} color information available. 
However, for this paper, we assume that these companions are not higher-mass companions as their \textit{Gaia} $G$ magnitudes are less than the other components. 
We also assume that they are not white dwarfs as, for those where the \textit{Gaia} resolved companion was detected, the magnitude differences in the red 832 nm filter from the speckle imaging are not as large as one might expect between a white dwarf and a low-mass star and a search on Simbad did not indicate the presence of a white dwarf.

Removing the six systems which include higher-mass or white dwarf components from our sample and Binary 46 as only one component of it was observed, we find 29 true wide binaries, nine speckle detected higher-order multiples, nine wide systems from \textit{Gaia} consisting of all low-mass stars, three Simbad higher-order multiples. 
For the remainder of this paper, we only consider these 50 systems where all components are low-mass K- and M-dwarfs.

\subsection{Examining the Properties of Low-mass Wide Binaries Observed by NESSI}
\begin{figure}
\centering
\includegraphics[width=0.7\textwidth]{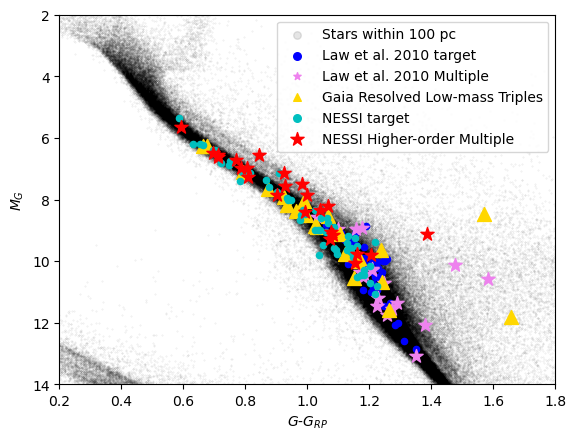}
\caption{\label{fig:nessihr}Color-magnitude diagram of our wide binary sample, with both the primary and secondary components shown. True wide binaries with no additional companions resolved by NESSI are marked by cyan circles, wide binaries containing a speckle-detected third companion are marked by red stars, and \textit{Gaia}-resolved higher-order multiple systems are shown as gold triangles. For the higher-order multiples detected by speckle imaging, both primary and secondary component are marked with stars ,however only one may be a binary. For the \textit{Gaia} resolved triples, only the two components observed by NESSI are shown and some of these have speckle detections themselves. We also include the systems observed by \citet{2010law} as the blue circles (true wide binaries - no detected third component) and violet stars (higher-order multiple).}
\end{figure}

With our sample of 50 low-mass systems, we can start to examine the higher-order multiplicity of low-mass wide binaries. 
Figure \ref{fig:nessihr} shows the color-magnitude diagram for the 50 systems with the speckle and \textit{Gaia} resolved systems highlighted. 
Overall, the higher-order multiplicity of our sample is $42.0\% \pm 10.9\%$ with the uncertainty calculated using Poisson statistics. 
This is in general agreement with the results of \citet{2020hartman} for K+K wide binaries found in the SUPERWIDE catalog. 
However, we will note that \citet{2020hartman} looked for over-luminous companions, which will find systems with different, but overlapping, separations and mass ratios than the current study. 
Some differences, therefore, are expected.

\begin{figure}
\centering
\includegraphics[width=0.7\textwidth]{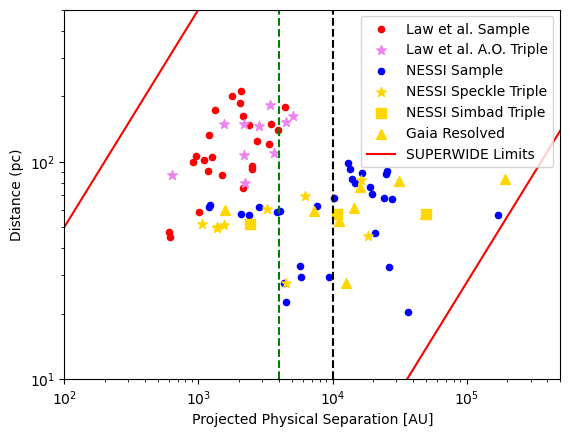}
\caption{\label{fig:nessiouter}Widest projected physical separations of the 50 systems in our sample plotted against distance to the primary component of the wide binary. Blue points show the location of systems which are true binaries according to NESSI and \textit{Gaia}. Gold points indicate the systems which are higher-order multiples according to NESSI (stars), \textit{Gaia} (triangles), and Simbad (squares). Red points and violet stars represent the same for the \citet{2010law} sample. The red lines indicates the upper (3600\arcsec) and lower (2\arcsec) search radii of SUPERWIDE in this plot. The vertical black dashed line shows the median dividing line used in our analysis using only our NESSI results while the vertical green dashed line represents the 4000 au upper selection used by \citet{2010law}.}
\end{figure}

We plot the parallax-estimated distance to the primary component against the projected physical separation of the 50 low-mass systems in Figure \ref{fig:nessiouter}, with blue circles representing true wide binaries, gold stars for speckle-detected multiple systems, gold squares for multiples with Simbad indications of multiplicity and gold triangles for \textit{Gaia} resolved triples. 
Red circles and violet stars represent systems from the \citet{2010law} sample and will be discussed later. 
For the \textit{Gaia} resolved triples, we used the widest binary separation in the system.
We note that some of the \textit{Gaia} resolved triples also are detected with speckle imaging, however, we mark all of the resolved triples the same. 

To examine whether we see a correlation between higher-order multiplicity fraction and projected physical separation, we split our sample at 10,000 au (roughly the median physical separation of our sample) and calculate these fractions separately. 
For those systems with widest separations less than 10,000 au, we get a fraction of $43.5\%\pm16.5\%$.
For widest separations greater than 10,000 au, the fraction falls to $40.7\%\pm14.6\%$. 
To compare the proportions of these two populations, we performed a hypothesis test that compares two independent population proportions.
Here the null hypothesis is $p_1 = p_2$, and, in order to reject this null hypothesis at a 5\% level of significance, $\lvert Z \rvert$ < 1.96, where the test statistic Z is defined as: $ Z = \frac{\hat{p}_1 - \hat{p}_2}{\sqrt{\hat{p}(1-\hat{p})(\frac{1}{n_1} - \frac{1}{n_2})}}$. 
Here, $\hat{p}$ is the combined proportion of binaries in our two samples ($42.0\%$, here),  $\hat{p_1}$ and $\hat{p_2}$ are the proportions of binaries in the two sample ($43.5\%$ and $40.7\%$, respectively), and $n_1$ and $n_2$ are the numbers in each sample (23 and 27, respectively). 
When comparing the proportion of binaries in these two separated ranges, we find that Z = 0.195, indicating that there is not sufficient evidence to conclude that there is a difference in the proportions of triples above and below 10,000 au. 
These results do not match the trend observed in solar-type wide binaries, where the higher-order multiplicity fraction is found to increase with separation.

One possibility for the absence of this trend in our results is that there are further currently unresolved companions in these wide systems. 
To investigate this, we examined the available \textit{Gaia} data for the stars in these systems as \textit{Gaia} contains several parameters which are known possible indicators of multiplicity. 
We stress that these parameters are indicators of multiplicity and should not be assumed to definitively say if a given star is actually an unresolved system or not. 
\citet{2023Tokovinin} looked at several of these parameters as a way to identify unresolved systems, RUWE, IPDfmp, and error in the radial velocity.
\citet{2023Tokovinin} found that they are able to detect 48\% of pairs identified with those three parameters using speckle imaging at SOAR. 
They also find that effectively all pairs with separations larger than 0.2\arcsec are resolved by IPDfmp, while pairs with large RUWE are less likely to be detected by speckle imaging, with the reasons being given that the companions are too close or too faint. 

\citet{2025cifuentes} also examined \textit{Gaia} multiplicity indicators in their study of M-dwarf multiplicity. 
Much like \citet{2023Tokovinin}, they used RUWE, IPDfmp, and radial\_velocity\_error. 
However, they expanded their selection to take into account other parameters as well. 
Their criteria are listed below:

\begin{itemize}
    \item RUWE $>2$
    \item ipd\_gof\_harmonic\_amplitude $>0.1$ \& RUWE $>1.4$
    \item ipd\_frac\_multi\_peak $>30$
    \item rv\_chisq\_pvalue $<0.01$ \& rv\_renormalized\_gof $>4$ \& rv\_nb\_transits $\geq 10$
    \item radial\_velocity\_error $\geq 10\SI{}{\kilo\metre\per\second}$
    \item duplicated\_source $= 1$ 
    \item non\_single\_star flagged
\end{itemize}

To examine if any components in our systems are flagged by \textit{Gaia}, we adopt a combination of the selection criteria described in \citet{2023Tokovinin} and \citet{2025cifuentes}. 
We use the majority of \citet{2025cifuentes} criteria with two exceptions. 
First, we lower the IPDfmp cut to 2 to match that of \citet{2023Tokovinin}.
Second, we remove the duplicated\_source check as \citet{2025cifuentes} stated that this was not a primary indicator of multiplicity.

Based on these criteria, we find that 27 of the 50 systems are flagged with potentially having third components. 
As with \citet{2023Tokovinin}, we find that even when \textit{Gaia} indicates a companion, there is a chance that it will remain undetected with speckle imaging. 
The RUWE $>2$ threshold flags seventeen systems as possible unresolved pairs but only ten are found in our sample, six are confirmed with our speckle measurements, and four are part of a resolved \textit{Gaia} triple. 
The IPDfmp $>2$ threshold also flags sixteen possible unresolved pairs but only eight are detected with the speckle observations and five are found as \textit{Gaia} resolved triples. 
Furthermore, we find six systems with NSS flags, with two in systems where there is a speckle imaging detection and two that are found as part of a \textit{Gaia} resolved triple.
One difference from \citet{2023Tokovinin} is that our observations do not recover all of the IPDfmp pairs, we find only thirteen of the sixteen systems which satisfies our IPDfmp cut.
We do note that when the selections for each \textit{Gaia} parameter are combined, all speckle detected and most of the resolved triple systems satisfy at least one of the criteria we put on the systems. 
For the \textit{Gaia} resolved systems that are not found, they are widely (at least $\rho>5\arcsec$) separated from each other.

If we count our speckle detected systems, \textit{Gaia} resolved wide triples, and systems where \textit{Gaia} indicates the presence of a companion, we get a higher-order multiplicity fraction of $62.0\%\pm14.2\%$.
We again examine how this fraction changes as a function of separation using 10,000 au again as a dividing line. 
For systems with the widest separations $<$10,000 au, we find that the fraction is $56.5\%\pm19.6\%$ and, for systems with widest separations $>$10,000 au, this fraction becomes $66.7\%\pm20.3\%$.
In the same manner as before, we test how similar these multiplicity fractions are. 
Plugging the updated values into the test statistic equation, we get a value for Z of -0.737, again not showing evidence for a statistical difference between the two multiplicity fractions.

\subsection{Comparison to the \citet{2010law} Sample}
One of the few papers to publish complete results for low-mass wide binaries, with both primary and secondary observed with high angular resolution imaging and results reported, was \citet{2010law}.
Using adaptive optics at Palomar and Keck, they observed 36 M+M wide binaries that were identified as part of the SLoWPoKES  survey \citep{2010dhital} and found a bias-corrected higher-order multiplicty fraction of $45\%\substack{+18\% \\ -16\%}$, a value that is marginally lower but still statistically consistent with our results.
\citet{2010law} also saw a bias-corrected rising trend in higher-order multiplicity as a function of widest physical separation, with systems below 2000 au having a fraction of $21\%\substack{+17\% \\ -7\%}$ and those greater than 4000 au having a fraction of $77\%\substack{+9\% \\ -22\%}$. 

For validation, we crossmatched the \citet{2010law} sample with \textit{Gaia} DR3 to obtain the source ids and other information. 
We then calculated the angular separation between the components using the \textit{Gaia} DR3 astrometry and recalculated the projected physical separations.
Additionally, we crossmatched this sample with SUPERWIDE and SIMBAD to check for resolved higher-order multiples.
All but one of the systems were found in SUPERWIDE, although some had low probabilities.
The one missing system, SLW1424+1856, was not in \textit{Gaia} DR2 and was therefore not included in SUPERWIDE.
This system is in \textit{Gaia} DR3 and is now included in our analysis. 
We also note that in \textit{Gaia} DR3, there is a third star nearby that has no astrometric solution and could potentially be the third component detected in the A.O. images. 
SLW 1548+0443 was marked as having a potential brown dwarf companion with a future paper being mentioned to confirm this \citep{2010law}. 
However, Simbad does not list any follow-up paper and thus we mark this system as a wide binary rather than triple. 
SLW1558+0231 was removed from the sample after it was discovered to be a higher-order multiple with a white dwarf companion (Gaia DR3 4412793818186163456). 
SLW2127-0040 was found to be a resolved triple with a higher-mass companion and was removed from our analysis.

In total, we find three \textit{Gaia} resolved triples, SLW1639+4226, SLW1424+1856, and SLW1357+1952. 
Two of these, SLW1639+4226 and SLW1424+1856, are listed in \textit{Gaia} DR3, but have one component with no astrometric solution or $G_{BP}-G_{RP}$ information.
However, these systems also have A.O. detections from \citet{2010law}, which could be the star resolved in \textit{Gaia}; We believe this is the case for SLW1639+4226 and SLW1424+1856. 
As with similar systems in the NESSI sample, we included these systems in this part of the analysis. 
A fourth system, SLW1439+5154, was determined to not be a higher-order multiple, due to the low probabilities of being a bound system with the other component from SUPERWIDE. 
This was only to the wide tertiary that was seen in SUPERWIDE, not SLW1439+5154 itself.
For the purpose of this paper, it is counted as a true wide binary.

Taking the 34 remaining systems from \citet{2010law}, Figure \ref{fig:nessihr} shows where they fall on the color-magnitude diagram and Figure \ref{fig:nessiouter} shows a plot of the projected physical separation vs. distance. 
The \citet{2010law} sample explores a different range of physical separations and distances, one which is closer in separation and farther away than our NESSI sample and goes to lower-mass M-dwarfs.
These differences are by design as we wanted a low-mass sample while also having a wider range of targets, hence the inclusion of K-dwarfs, we wanted a sample that took advantage of the high resolution capabilities of NESSI, and we wanted a sample that overlapped and expanded upon the results from \citet{2010law}. 
10 of the 34 systems have either an adaptive optics detection or are a resolved triple in \textit{Gaia}. 
Combining this sample with our NESSI sample, we find 31 higher-order multiples out of 84 total systems, giving a higher-order multiplicity fraction of $36.9\%\pm7.8\%$. 
This value is consistent with our results using just the NESSI data. The inclusion of \citet{2010law} improves the statistics in our analysis of the higher-order multiplicity fraction as a function of separation.
If we set our dividing line at 4,000 au (matching where \citet{2010law} set their upper selection limit), we find that for systems with separations below this cutoff, the higher-order multiplicity fraction is $34.1\%\pm10.1\%$, while for systems with separations greater than our cutoff, it is $40.0\%\pm11.8\%$.
If we examined the test statistic Z for this comparison, we find that our value for Z is -0.561, again saying that the two fractions are not significantly different.

\begin{figure}
\centering
\includegraphics[width=0.7\textwidth]{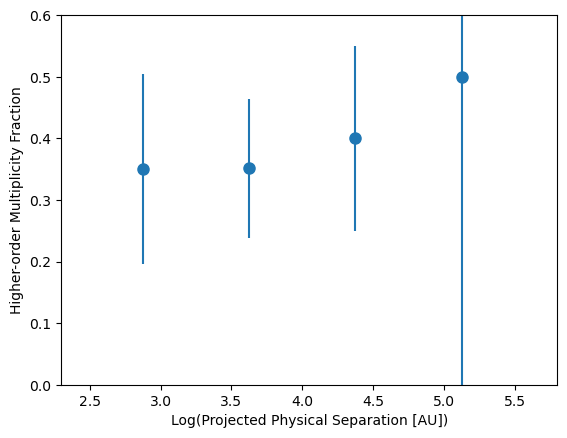}
\caption{\label{fig:homfvphysep}Higher-order multiplicity fraction as a function of widest projected physical separation in the combined sample of \citet{2010law} and our NESSI observations. The final bin at $10^5$ contains only 2 systems and, therefore, its value is questionable. We do not see evidence of increasing higher-order multiplicity at increasing physical separations.}
\end{figure}

We break this down more in Figure \ref{fig:homfvphysep}, where we we plot the higher-order multiplicity fraction over 4 bins with the error bars representing Poisson uncertainty on the values. 
This does not include binaries where \textit{Gaia} indicators such as RUWE and IPDfmp point to an unresolved binary. 
As shown, we do not see an increase using only the high-resolution imaging and literature results. 
We are skeptical of the apparent drastic increase in higher-order multiplicity fraction beyond $10^{4.5}$ au as the last bin only contains 2 systems, but the lack of an increasing trend is significant.  

\begin{figure}
\centering
\includegraphics[width=0.7\textwidth]{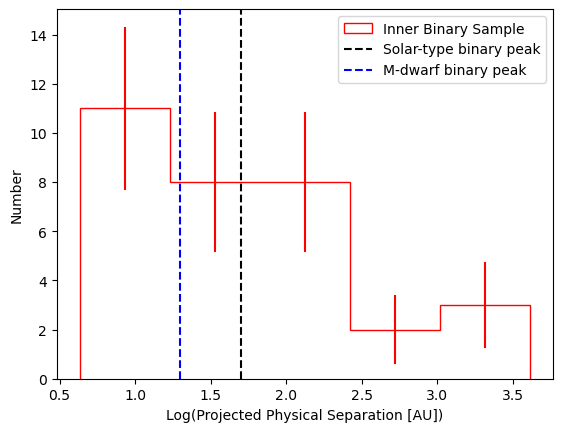}
\caption{\label{fig:inner}Distribution of the projected physical separations for the 32 inner binaries that were detected in either our NESSI search, \citet{2010law}, or \textit{Gaia} DR3 as resolved triples. The two vertical dashed lines represent the peaks of the separation distribution for solar-type stars at 51 au (black line, from \citet{2010raghavan}) and for M-dwarfs at 20 au (blue line, from \citet{2019winters}). We note the small number statistics here and the need to find more of these close binaries in very wide systems to fill in this distribution.}
\end{figure}

With the additional stars from \citet{2010law}, we construct a histogram of the projected physical separations of the inner binaries that were detected through either the high resolution imaging or \textit{Gaia} searches (Figure \ref{fig:inner}). 
In total, 32 binaries are included on this plot.
Previously, our analysis has been on a system basis. 
For this part, we expand this to include the separations for close companions in quadruple systems, which were not taken into account previously. 
In these cases, we add in the results from the speckle observations when they add a new companion \textit{Gaia} did not detect. 
There are four such cases. 
These are from \textit{Gaia} resolved triple systems, however, there is one system from \citet{2010law}, SLW2456+0017. 
For speckle detected systems, the separation found from the red filter is used for the separation of the inner binary. 
The vertical lines represent the peaks of the binary distributions for solar-type binaries - 51 au as suggested by \citet{2010raghavan} - and M-dwarf binaries - 20 au as suggested in \citet{2019winters}.
There is a potential increase in companions detected at lower separations compared to the peaks of the solar-type and M-dwarf distributions, which may suggest that these systems might have undergone angular momentum transfer at some point in their lives to form closer binaries. 
However, more observations of low-mass triple systems are needed to examine this further.

If we include \textit{Gaia} indicators like in Section 3.5, we do get a larger higher-order multiplicity value and possibly begin to see the trend of higher-order multiplicity with projected physical separation. 
Overall, the fraction increases to $53.6\%\pm9.9\%$. 
When examining this as a function of physical separation in a similar manner to above, we find a fraction of $45.5\%\pm12.3\%$ for those systems with projected physical separations below 4,000 au and $62.5\%\pm13.7\%$ for those above this threshold. 
Although this suggests an increase in the multiplicity fraction, the uncertainties remain too high to be definitive, and it is clear that a much larger census will ultimately be required to answer the question. 
This is supported by examining the test statistic Z again. Here, our value for Z is -1.56. 
While it is slightly larger than the previous values, it is not large enough for us to confidently say that there is a difference in these two proportions at a significance level of 5\%.
Therefore, our results currently indicate unresolved binary companions cannot account for the lack of an observed increase in higher-order multiplicity fraction with separation.
However, more observations are needed to confirm this as we did see an increase in the Z value as potential unresolved binary candidates were added in.

\subsection{Biases in Our Sample}
There are several biases that are not accounted for in this analysis. 
The first one is that \textit{Gaia} is biased against detecting wide binaries where one component may be a binary itself.
It has been reported that the presence of an unresolved or partially resolved companion can have an effect on the \textit{Gaia} astrometric solution \citep{2023Tokovinin}. 
This can vary from the star having a higher error on the parallax to the entire system only having sky coordinates (R.A., Decl.) and no parallax or proper motion data \citep{2023Medan}. 
The former might mean that the binary probability in SUPERWIDE would have fallen below our selection criteria for this sample because the parallaxes of the components would have been inconsistent (one being off from the real value) while the latter would mean that system would have been overlooked by the SUPERWIDE survey.
SUPERWIDE also used a quality cut excluding stars with parallax errors larger than 15\%, which would have excluded the inner pairs of triple systems with large astrometric uncertainties.
The effect of an unresolved companion on \textit{Gaia} data is beyond the scope of this paper and we do not here attempt to try and make a correction for it.  
Furthermore, more stars have had their astrometric information either updated or added in between \textit{Gaia} DR2 and DR3, superseding some of the information used to build SUPERWIDE. 

To try to quantify how this affects the completeness of the SUPERWIDE census, we apply the method of \citet{2023Tokovinin} to independently select a sample of wide systems from the \textit{Gaia} Catalog of Nearby Stars (GCNS, \citet{2021gcns}). 
See Section \ref{sec:bin25} for the requirements.
We modify the second condition slightly and adopt a lower cut on separation by requiring that the angular separations between the components be larger than 2\arcsec to match that of the SUPERWIDE catalog, and we set the upper limit to 100,000 au to grab wider pairs. 
We also select only stars that are within 100 pc and have proper motions $>$40 mas/yr, the former to match our distance cut for this study and the latter to match that of SUPERWIDE. 
We apply these conditions to the GCNS and find a sample of 17,198 systems.
Crossmatching this with SUPERWIDE, we recover 14,720 systems, which gives an 85.6\% match. 

Looking closer at the systems that do not match, we find 2007 wide binary systems and 471 higher-order multiples. 
We do not attempt to exclude nearby clusters and moving groups, which means some of the higher-order multiples could be cluster members, not bound systems.
As mentioned above, there are several reasons why these systems do not appear in the SUPERWIDE sample. 
Some now have updated astrometric information that only now allows them to be matched to nearby stars. 
This is the case for several wide systems in SUPERWIDE, where a third/fourth additional component can now be matched to a previously found pair/triple. 
This is counted as not a match in the above cross-match as all members are not in SUPERWIDE. 
Some of the additional wide binary systems also have significant proper motion differences between the two components that would have excluded them from the SUPERWIDE selection algorithm, but are identified using the relaxed conditions in \citet{2023Tokovinin}. 
These verifications, however, indicate that while there are likely legitimate systems missing in SUPERWIDE, our survey is still relatively complete. 

We also know we are not surveying the entire separation-contrast space for these systems. 
The diffraction limit for NESSI on the WIYN telescope is roughly 40 mas, which is about 4 au for a star at 100 pc. 
From \citet{2019winters}, it is known that M-dwarf binaries in the field have a separation distribution that peaks between 4-20 au. 
Thus, the low-mass stars have a shorter distribution of orbital separations compared with solar-type binaries, which peak at 51 au \citep{2010raghavan}.
K-dwarf binaries will likely have a peak distribution that is between the two.
Our sample has been designed to select wide binaries consisting of K- and M-dwarfs, therefore close companions will most likely be K- and M-dwarfs themselves. 
If the separation distribution of close companions in these higher-order multiples mirrors the field distribution \citep{2014tokovininb}, then we should be finding a large fraction of potential K- and M-dwarf close companions.
This is not taking into account potential effects the outer component in the system might exert on the close binary. 
However, we still know that there are binaries in the field with separations and contrasts below what we can detect. 
To partially account for this, we include known spectroscopic and eclipsing binaries in the literature as noted above, however, this list is incomplete. 
To fully examine these systems, further spectroscopic follow-up is needed to search these systems for close companions.

\section{\textbf{Discussion}}

An examination of previous wide binary surveys shows that roughly half of solar-type wide binaries are actually higher-order multiples \citep{2010raghavan, 2014tokovinina, 2014tokovininb}. 
Evidence also suggests that the fraction of systems that are higher-order multiples increases with the physical separation of the widest components \citep{2010law, 2010raghavan, 2014tokovinina, 2014tokovininb}. 
Intuitively this makes sense, as pointed out in \citet{2017tokovinin}, because there is simply more space in a wider system where a third component could reside and also because higher-order systems tend to have more mass which gives the system a higher binding energy allowing them to survive longer as bound systems. 
Because of these findings, many theories and models regarding wide binary formation have taken this large higher-order multiplicity fraction into account. 
However, these findings have been mostly based on studies involving binaries with a solar-type or higher-mass primary star.
In one of the few studies to examine low-mass stars, \citet{2010law} seemed to find that these trends held true for low-mass binary systems as well.
However, their sample was limited in number (36) and in separation range (out to roughly 6500 au), which left the high frequency of multiples in wide low-mass systems an open question \citep{2023offner}. 

Our study expands upon this survey in both areas by more than doubling the number of systems (50 systems added) and also by increasing the wide binary separations examined (out to $>$30,000 au). 
Our result of 42.0$\%\pm$10.9$\%$ is consistent with the higher-order multiplicity fraction of \citet{2010law}, $45\%\substack{+18\% \\ -16\%}$, but improves the statistical confidence. 
However, we do not see concrete evidence of a trend towards higher multiplicity with increasing physical separation, even if we include potential indicators of multiplicity from \textit{Gaia}. 
This mirrors results seen in \citet{2022hartman} for a sample of overluminous K+K systems, with the higher-order fraction remaining flat with orbital separation. 
One possibility for why we do not see the rise suggested by \citet{2010law} is that we do not bias correct our results. 
In \citet{2010law}, they modeled how the presence of an unresolved companion would effect their selections and found that binaries with unresolved companions would be suppressed by factors of 1.4 to 3.1.
However, \citet{2010law} used photometric distances for their analysis, which are more sensitive to the presence of an unresolved companion than we believe our current \textit{Gaia}-based selection is.
Another possibility for why we do not see the rise is that the companions are too close to detect with speckle imaging. 
A more stringent test of this requires a more complete search for close companions with spectroscopic follow-up.

Our current results join the mounting evidence that wide binary formation is most likely not dominated by a single mechanism and is likely a combination of factors.
In \citet{2022hwang}, they examined the eccentricity distribution for a sample of binaries selected from \textit{Gaia} and found that, for systems with separations $>$1000 au, the distribution is superthermal, meaning that the power law term on the eccentricity distribution function is greater than 1. 
This is expected if the binaries formed through turbulent fragmentation \citep{2014bate} and/or the dynamical unfolding of compact triples \citep{2012reipurth}, both of which have highly eccentric binaries at larger separations.
However, \citet{2022hwang} also found that their distribution of eccentricities for binaries with separations $10^3$ au to $10^{4.5}$ au is constant rather than increasing as is expected for binaries formed through the two previously mentioned methods. 
They mention that this could be due to a number of factors, including systematics in their analysis along with gravitational interactions between passing stars and other bodies in the Galaxy.

For our study, we believe our results point to dynamical unfolding not being the main driver of wide binary formation over the separation range we examine.
If the dynamical unfolding of compact triples was dominant, one would expect all or nearly all of the systems to be higher-order multiples \citep{2012reipurth}.
Our results do not show this and are, overall, lower than previous results for solar-type and higher-mass binaries. 
This may be evidence that the higher-order multiplicity fraction of wide binaries is influenced by the individual multiplicity fractions of its components, which would happen if turbulent fragmentation or the binding of adjacent cores were prominent scenarios in wide binary formation \citep{2014bate,2014tokovininb,2017tokovinin}. 
However, if dynamical unfolding was the dominant wide binary formation scenario, one would expect the inner component's separation distribution to be skewed to closer separations due to the interactions between components \citep{2012reipurth}. 
Our distribution of inner binary separations (Figure \ref{fig:inner}) does appear to be skewed towards smaller separations given that this distribution is still rising at roughly 10 au, below the peaks in the separation distributions of solar-type and M-dwarf binaries from previous papers \citep{2010raghavan,2019winters}. 
This implies that some dynamical interactions may have occurred in these systems if the systems initially formed from the same parent distribution as put forth in \citet{2014tokovininb}. 
More work needs to be done in this area to increase the statistical confidence in the higher-order multiplicity fractions.
We are currently obtaining observations of more low-mass wide binary systems using the Quad-camera Wavefront-sensing Six-channel Speckle Interferometer (QWSSI, \citet{2020clark}). 
Furthermore, searching for closer binaries using radial velocity measurements or mining wide-field photometric surveys like TESS will allow us to search nearly the entire separation range. 
These observations, combined with long term orbital monitoring, will allow us to fully characterize these systems.

\section{\textbf{Conclusions}}

We present our analysis on the individual components of 56 wide systems and one component of another system using NESSI at the WIYN telescope. 
We find 29 true wide binaries, nine speckle detected systems, nine \textit{Gaia} resolved triples, three higher-order multiples found in previous surveys, and 6 higher-order multiples where one of the additional components is of higher-mass than a K- or M-dwarf or a white dwarf. 
These systems span from 1000 to $>$100,000 au and comprise of only K- and M-dwarfs.
With just the sample of 50 low-mass systems, we determine a higher-order multiplicity fraction of $42.0\%\pm10.9\%$, which is lower than previous estimates for solar-type systems. 
This value is statistically consistent with, but improves upon, the higher-order fraction estimate by \citet{2010law}, one of the few studies of low-mass wide binaries. 
However, we do not see the trend of increasing higher-order multiplicity with increasing physical separation that was suggested in this previous study.
If \textit{Gaia} indicators of multiplicity are added to a sample comprised of both the \citet{2010law} sample and our sample, our derived higher-order multiplicity fractions are constant as a function of projected physical separation.
This points towards the dynamical unfolding of triple systems not being the dominant formation mechanism for these low-mass wide binaries as one would expect nearly all of the wide binaries to be higher-order multiples if this was the case.
We also examined the distribution of the inner binaries in both our sample and that of \citet{2010law} and find that the distribution appears to peak at a lower separation than the field population of solar-type or M-dwarf binaries. 
This may point to some dynamical interactions between components, which would push the inner binaries into closer orbits and indicates dynamical unfolding may play a role in low-mass wide binary evolution, but maybe not the dominant mechanism in their formation.

\section{Acknowledgments}.

We would like to thank the anonymous referee for all the helpful and insightful comments. Z.D.H was funded for this work through NN-EXPLORE Grant (Grant 1623647). Z.D.H. would like to thank Catherine Clark, Jos\'{e} A. Caballero, Julio Chanam\'{e}, Sebasti\'{a}n Javier Vilaza Dallago, Todd Henry, Douglas Gies, Xiaochun He, Bokyoung Kim, Maxwell Moe, and Andrei Tokovinin for their insightful comments and help for this project. 

This work has made use of data from the European Space Agency (ESA) mission {\it Gaia} (\url{https://www.cosmos.esa.int/gaia}), processed by the {\it Gaia}
Data Processing and Analysis Consortium (DPAC,\url{https://www.cosmos.esa.int/web/gaia/dpac/consortium}).
Funding for the DPAC has been provided by national institutions, in particular the institutions participating in the {\it Gaia} Multilateral Agreement.

Based on observations at NSF Kitt Peak National Observatory, NSF NOIRLab (NOIRLab Prop. ID 2019A-0301; PI: Z, Hartman), which is managed by the Association of Universities for Research in Astronomy (AURA) under a cooperative agreement with the U.S. National Science Foundation. The authors are honored to be permitted to conduct astronomical research on I'oligam Du’ag (Kitt Peak), a mountain with particular significance to the Tohono O’odham.

Some of the observations in the paper made use of the NN-EXPLORE Exoplanet and Stellar Speckle Imager (NESSI). NESSI was funded by the NASA Exoplanet Exploration Program and the NASA Ames Research Center. NESSI was built at the Ames Research Center by Steve B. Howell, Nic Scott, Elliott P. Horch, and Emmett Quigley.

Data presented herein were obtained at the WIYN Observatory, or the CTIO SMARTS 1.5m, or MINERVAAustralis from telescope time allocated to NN-EXPLORE through the scientific partnership of the National Aeronautics and Space Administration, the National Science Foundation, and the NOIRLab. This work has made use of data from the European Space Agency (ESA) mission {\it Gaia} (\url{https://www.cosmos.esa.int/gaia}), processed by the {\it Gaia} Data Processing and Analysis Consortium (DPAC, \url{https://www.cosmos.esa.int/web/gaia/dpac/consortium}). Funding for the DPAC has been provided by national institutions, in particular the institutions participating in the {\it Gaia} Multilateral Agreement.

This work has made use of Python and various python packages and services, including NumPy \citep{numpy}, SciPy \citep{2020SciPy-NMeth}, Matplotlib \citep{huntermatplotlib}, Astropy \citep{2013astropy,2018astropy2}, and Astroquery \citep{2019astroquery}. This research has made use of NASA's Astrophysics Data System.

\bibliography{sample631}{}
\bibliographystyle{aasjournal}

\begin{deluxetable}{cccccccccc}
\tablewidth{700pt}
\tabletypesize{\scriptsize}
\tablehead{
\colhead{Binary ID} & \colhead{R.A.} & \colhead{Decl.} & \colhead{$\mu_{R.A}$} & \colhead{$\mu_{Decl.}$} & \colhead{$\pi$} & \colhead{$G$}& \colhead{$G_{BP} - G_{RP}$} & \colhead{\textit{Gaia} DR3 ID} & \colhead{Observation Date} \\
\colhead{} & \colhead{[\degree]} & \colhead{[\degree]} & \colhead{[mas/yr]} & \colhead{[mas/yr]} & \colhead{[mas]} & \colhead{[mag.]} & \colhead{[mag.]} & \colhead{} & \colhead{2019}}
\tablecaption{\label{tab:targets}\textit{Gaia} Information on Observed Wide System Components}
\startdata
1 & 290.542 & 70.846 & 38.833 & -67.688 & 16.531 & 11.768 & 1.735 & 2262681645906587904 & 6/18 \\
1 & 290.547 & 70.831 & 36.111 & -65.359 & 16.382 & 13.711 & 2.417 & 2262681547123067392 & 6/18 \\
2 & 243.069 & 41.243 & 90.624 & -118.739 & 19.409 & 10.057 & 1.244 & 1381579138175507840 & 6/18 \\
2 & 243.07 & 41.23 & 93.713 & -119.491 & 19.38 & 13.38 & 2.726 & 1381579142471217536 & 6/18 \\
3 & 12.616 & 59.502 & 162.692 & -159.997 & 21.355 & 9.812 & 1.241 & 426757197608443904 & 6/19 \\
3 & 12.59 & 59.388 & 162.613 & -157.537 & 21.367 & 12.845 & 2.383 & 426732767830735744 & 6/19 \\
4 & 23.918 & 76.002 & 283.031 & 53.769 & 12.599 & 9.838 & 1.004 & 559852564744528384 & 7/25 \\
4 & 23.964 & 76.053 & 283.045 & 53.847 & 12.582 & 13.691 & 2.443 & 560228047964634624 & 7/25 \\
5 & 333.66 & -10.327 & 103.293 & -109.813 & 10.855 & 11.057 & 1.169 & 2615507406771731328 & 7/15 \\
5 & 333.646 & -10.288 & 103.619 & -110.1 & 10.971 & 13.401 & 2.199 & 2615507716009382528 & 7/15 \\
6 & 200.104 & -1.659 & 129.267 & -253.466 & 44.391 & 10.741 & 2.132 & 3686261307923490688 & 6/19 \\
6 & 200.053 & -1.679 & 130.873 & -253.227 & 44.466 & 12.262 & 2.538 & 3686259727375524864 & 6/19 \\
7 & 244.931 & 3.016 & -101.811 & -167.87 & 11.283 & 10.92 & 1.094 & 4436102536989157760 & 7/20 \\
7 & 244.908 & 2.969 & -102.237 & -166.835 & 11.421 & 14.366 & 2.294 & 4436101815434645632 & 7/20 \\
8 & 349.395 & 6.39 & 170.439 & -249.487 & 48.918 & 11.429 & 2.49 & 2664156432614468864 & 6/18 \\
8 & 349.036 & 6.742 & 173.642 & -250.616 & 48.976 & 11.994 & 2.631 & 2664368290465455232 & 6/18 \\
9 & 225.687 & 18.171 & 162.034 & -209.185 & 11.095 & 10.977 & 1.149 & 1188563037310217216 & 7/13 \\
9 & 225.75 & 18.222 & 161.34 & -209.45 & 10.92 & 14.32 & 2.5 & 1188586779889479936 & 7/13 \\
10 & 214.807 & 24.332 & -353.818 & 127.652 & 19.922 & 11.053 & 1.746 & 1255695162853191936 & 6/18 \\
10 & 214.815 & 24.33 & -365.413 & 126.742 & 19.801 & 12.802 & 2.241 & 1255695261636767488 & 6/18 
\enddata
\tablecomments{See full table online.}
\end{deluxetable}

\begin{deluxetable}{cccccccc}
\tablewidth{700pt}
\tabletypesize{\scriptsize}
\rotate
\tablehead{
\colhead{Binary ID} & \colhead{WDS ID} & \colhead{\textit{Gaia} DR3 $\mathrm{ID}_1$} & \colhead{Source} & \colhead{Is 1 Binary?} & \colhead{\textit{Gaia} DR3 $\mathrm{ID}_2$} & \colhead{Source}& \colhead{Is 2 Binary?}}
\tablecaption{\label{tab:nessisimbad}Literature Search Results for Close Companions}
\startdata
1 & 19222+7051 & 2262681645906587904 & -- & N & 2262681547123067392 & -- & N \\
2 & -- & 1381579138175507840 & TS -- \citet{2018collins} & N & 1381579142471217536 & TS -- \citet{hartman2011} & Y \\
3 & 00505+5930 & 426757197608443904 & PMA -- \citet{2021brandt} & N & 426732767830735744 & -- & N \\
4 & 01357+7600 & 559852564744528384 & PMA -- \citet{2021brandt} & N & 560228047964634624 & -- & N \\
5 & 22147-1019 & 2615507406771731328 & -- & N & 2615507716009382528 & -- & N \\
6 & 13203-0140 & 3686261307923490688 & PMA -- \citet{2021brandt} & N & 3686259727375524864 & -- & N \\
7 & 16197+0301 & 4436102536989157760 & RV -- \citet{2009sozzetti} & N & 4436101815434645632 & HRI -- \citet{2020lamman} & N \\
8 & -- & 2664156432614468864 & HRI -- \citet{2013jodar} & N & 2664368290465455232 & -- & N \\
9 & 15027+1809 & 1188563037310217216 & PMA -- \citet{2021brandt} & N & 1188586779889479936 & HRI -- \citet{2020lamman} & N \\
10 & 14193+2421 & 1255695162853191936 & PMA -- \citet{2021brandt} & Y & 1255695261636767488 & -- & N 
\enddata
\tablecomments{See full table online.}
\end{deluxetable}

\begin{deluxetable}{cccccccccccccccccccccc}
\tablewidth{700pt}
\tabletypesize{\tiny}
\rotate
\tablehead{
\colhead{Binary ID} & \colhead{\textit{Gaia} DR3 $\mathrm{ID}_1$} & \colhead{$\Delta m_{b}$} & \colhead{$\theta_{b}$} & \colhead{$\rho_{b}$} & \colhead{$\Delta m_{r}$} & \colhead{$\theta_{r}$} & \colhead{$\rho_{r}$} & \colhead{\textit{Gaia} DR3 $\mathrm{ID}_2$} & \colhead{$\Delta m_{blue}$} & \colhead{$\theta_{blue}$} & \colhead{$\rho_{blue}$} & \colhead{$\Delta m_{red}$} & \colhead{$\theta_{red}$} & \colhead{$\rho_{red}$} & \colhead{\textit{Gaia} Triple} & \colhead{$P_{Wide}$} & \colhead{$\rho_{Wide}$} \\
\colhead{} & \colhead{} & \colhead{[mag.]} & \colhead{[\degree]} & \colhead{[\arcsec]} & \colhead{[mag.]} & \colhead{[\degree]} & \colhead{[\arcsec]} & \colhead{} & \colhead{[mag.]} & \colhead{[\degree]} & \colhead{[\arcsec]} & \colhead{[mag.]} & \colhead{[\degree]} & \colhead{[\arcsec]} & \colhead{} & \colhead{[\%]} & \colhead{[au.]}}
\tablecaption{\label{tab:nessiresults}NESSI Wide Binary Observations and Results.}
\startdata
1 & 2262681645906587904 & -- & -- & -- & -- & -- & -- & 2262681547123067392 & -- & -- & -- & 2.75 & 186.942 & 0.501 & N & 99.184 & 3254.89 \\
2 & 1381579138175507840 & -- & -- & -- & -- & -- & -- & 1381579142471217536 & -- & -- & -- & -- & -- & -- & N & 99.995 & 2409.77 \\
3 & 426757197608443904 & -- & -- & -- & -- & -- & -- & 426732767830735744 & -- & -- & -- & -- & -- & -- & Y & 99.995 & 19716.48 \\
4 & 559852564744528384 & -- & -- & -- & -- & -- & -- & 560228047964634624 & -- & -- & -- & -- & -- & -- & N & 99.995 & 14824.83 \\
5 & 2615507406771731328 & -- & -- & -- & -- & -- & -- & 2615507716009382528 & -- & -- & -- & -- & -- & -- & N & 99.992 & 13598.47 \\
6 & 3686261307923490688 & -- & -- & -- & -- & -- & -- & 3686259727375524864 & -- & -- & -- & -- & -- & -- & N & 99.995 & 4501.5 \\
7 & 4436102536989157760 & -- & -- & -- & -- & -- & -- & 4436101815434645632 & -- & -- & -- & -- & -- & -- & N & 99.995 & 16701.31 \\
8 & 2664156432614468864 & -- & -- & -- & -- & -- & -- & 2664368290465455232 & -- & -- & -- & -- & -- & -- & N & 99.995 & 36866.12 \\
9 & 1188563037310217216 & -- & -- & -- & -- & -- & -- & 1188586779889479936 & -- & -- & -- & -- & -- & -- & N & 99.994 & 25450.59 \\
10 & 1255695162853191936 & 2.67 & 176.106 & 0.244 & 1.96 & 174.688 & 0.226 & 1255695261636767488 & -- & -- & -- & -- & -- & -- & N & 99.995 & 1364.22 
\enddata
\tablecomments{See full table online.}
\end{deluxetable}

\begin{deluxetable}{cccccc}
\tablewidth{700pt}
\tabletypesize{\scriptsize}
\tablehead{
\colhead{Binary ID} & \colhead{$\mathrm{ID}_1$} & \colhead{$\mathrm{ID}_2$} & \colhead{$\mathrm{ID}_3$} & \colhead{$\mathrm{ID}_4$} & \colhead{Low-mass Triple}}
\tablecaption{\label{tab:nessitriple}NESSI Triple Systems from SUPERWIDE with Components Observed during NESSI Campaign.}
\startdata
3 & 426732767830735744 & 426757197608443904 & 426732767830736768 & -- & N\\
12 & 4518377377198705280 & 4518283162795992320 & 4518377377198705536 & -- &Y\\
13 & 1987197396963763712 & 1987983410339434624 & 1987197396959607936&-- &Y\\
14 & 1632112552014817024 & 1632110529080430592 & 1632110322923341184&-- &Y\\
17 & 1609314934323173120 & 1609405545248504576 & 1609314938618876416 & 1609405545248504192 & N\\
22 & 2741792433256603904 & 2741790990147593344 & 2741790990147593216 & -- &Y\\
29 & 389658300887082496 & 388149427335376256 & 389658300887082624 & -- &Y\\
30 & 1727367612307674368 & 1727367509227099776 & 1727555834952561664&-- &N\\
33 & 2829469414599661568 & 2829469414599661440 & 2829468727404896000&-- &N\\
34 & 1318430994232581504 & 1318431062952058752 & 1318430989938628992 & -- &Y\\
38 & 1662678945197952768 & 1662678945198351616 & 1662676467001283584&-- &N\\
43 & 4370790721670081920 & 4370790721659321344 & 4370790103191277568&-- &Y\\
45 & 6899611699689482880 & 6899592733113659392 & 6899611699689482496&-- &N\\
48 & 3920450962448498816 & 3920498104008416768 & 3920498104009539456&-- &Y\\
55 & 4331385186994355072 & 4331478508043788032 & 4331385186992338304 & -- & Y
\enddata
\end{deluxetable}

\end{document}